\documentclass[twocolumn,prb,citeautoscript,superscriptaddress]{revtex4}
\usepackage{amsfonts,amsmath,amssymb}
\usepackage[dvips]{graphicx,color}
\input epsf 
\newcommand{\fig}{Figs}

\begin{document}

\title{Pores in Bilayer Membranes of Amphiphilic Molecules: 
Coarse-Grained Molecular Dynamics Simulations Compared with Simple Mesoscopic Models.} 

\author{C. Loison}
\email{loison@cpfs.mpg.de}
\affiliation{Max Planck Institut f\"ur Chemische Physik fester Stoffe, 
N\"othnitzer Str. 40, D-01187 Dresden, Germany}

\author{M. Mareschal}
\affiliation{Centre Europ\'een de Calcul Atomique et Mol\'eculaire,
ENS Lyon, 46, All\'ee d'Italie, F-69007 Lyon, France}

\author{F. Schmid}
\email{schmid@physik.uni-bielefeld.de}
\affiliation{Fakult\"at f\"ur Physik, Universit\"at Bielefeld,
Universit\"atsstra{\ss}e 25, D-33615 Bielefeld, Germany}

\date{\today}
\begin{abstract}
\noindent
We investigate pores in fluid membranes by molecular dynamics simulations of an amphiphile-solvent mixture, using a molecular coarse-grained model. The amphiphilic membranes self-assemble into a lamellar stack of amphiphilic bilayers separated by solvent layers. We focus on the particular case of  tensionless membranes, in which pores spontaneously appear because of thermal fluctuations. Their spatial distribution is similar to that of a random set of repulsive hard discs. The size and shape distribution of individual pores can be described satisfactorily by a simple mesoscopic model, which accounts only for a pore independent core energy and a line tension penalty at the pore edges. In particular, the pores are not circular: their shapes are fractal and have the same characteristics as those of two dimensional ring polymers. Finally, we study the size-fluctuation dynamics of the pores, and compare the time evolution of their 
contour length to  a random walk in a linear potential.
\end{abstract}
\maketitle

\section{Introduction}

Fluid lipid bilayers are the basic material of biological membranes. Pores in such bilayers play an important role in  the diffusion of small molecules across biomembranes \cite{Freeman_BJ_94,Lawaczeck_BBPC_88,Paula_BJ_96}. In the last decade, the interest in pore formation in bilayer membranes has greatly increased with the development of electroporation -- in this technique, an intense electric field is applied for a short time allowing bulky hydrophilic molecules to permeate through the lipid membranes of a cell or a vesicle.  Additionally, the formation of pores in bilayers is supposed to be one key step of the fusion of membranes \cite{Safran_BJ_01,Hed_BJ_03,Muller_JCP_02}. 

One difficulty in studying the pores experimentally  is that they are
 usually not visible in optical microscopy. Yet, they have been investigated  using indirect techniques 
such as permeation measurements with micropipettes~( \textit{e.g.} Ref. ~~\onlinecite{Olbrich_BJ_00})
 or small angle neutron scattering~\cite{Holmes_JPF_93}.
The mechanisms of permeation through a membrane obviously depend on the size and the life-time distributions of the pores. Such data are difficult to obtain experimentally because of the small life-time and  the small dimensions of the pores (less than a millisecond and a few nanometers~\cite{Zhelev_BBA_93,Melikov_BJ_01}). Therefore, numerical calculations -- either with molecular models~\cite{Marrink_BJ_96,Mueller_JCP_96,Holyst_JCP_97,Marrink_JACS_01,Zahn_CPL_02} or with density functional theories~\cite{Netz_PRE_96,Talanquer_JCP_03} -- have proven useful to study the local structure of  defects in 
amphiphilic bilayers and lamellar phases.

\begin{figure}[h!]
\begin{center}
\includegraphics[height=7cm,angle=-90]{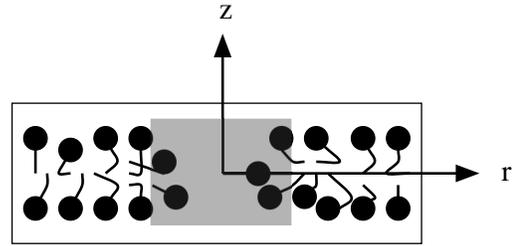}
\caption{Schematic cartoon of a hydrophilic pore in an amphiphilic bilayer (inspired by Ref.~~\onlinecite{Glaser_BBA_88}). 
The dark discs are the hydrophilic heads of the amphiphiles, the thin lines their hydrophobic tails. 
The solvent above/under the bilayer and in the pore is not represented. 
In the grey region, the structure of the bilayer is perturbed.}
\label{hole_types}
\end{center}
\end{figure}

Models for pores dynamics\cite{Freeman_BJ_94,Sung_BJ_97,Shillcock_BJ_1998, Shillcock_BJ_96} are
often based on Glaser's model of pore formation \cite{Glaser_BBA_88}. He distinguishes between three stages. In the first stage, the bilayer thickness fluctuates because of thermal fluctuations. Some hydrophobic tails happen to be exposed to the solvent. In the second stage, the solvent spans through the hydrophobic layer, creating a hydrophobic pore. If this pore expands, it becomes energetically favorable for the amphiphiles in the edge of the pore to reorient. Finally, in the third stage, the pore possesses a hydrophilic edge (see Fig.~\ref{hole_types}).
The final hydrophilic pores carry an energy  $E$ 
that depends on the contour length  $c$ and the area  $a$ of the pore.
Lister \cite{Lister_PL_75} suggested that 
\begin{equation}
\label{energy_pore}
E=E_0+\lambda c -\gamma a,
\end{equation}
 where $E_0$ is equivalent to a core energy, $\lambda$  is the line tension of the pore edge, 
and $\gamma$ the surface tension of the bilayer. 
In most cases, $\lambda >0$ and $\gamma \geq 0$.
According to Lister's model [Eq.(1)], 
for strictly positive surface tension, 
pores larger than a critical size are stable and the membrane may break. 
For tensionless membranes ($\gamma=0$, the case we investigate here)
the energy increases monotonously when the pore grows. 
Pores of any size are unstable, and the membrane is stable.

The aim of this article is to compare the results of molecular dynamics simulations to 
simple mesoscopic models of pores in bilayers based on Eq.~(\ref{energy_pore}).
 The comparison has turned out to be fruitful for various models, 
and  permits to bridge the gap between the descriptions at different  
length and time scales.
M\"uller and coworkers \cite{Mueller_JCP_96} could show that pore formation in 
stretched bilayers composed of diblock-copolymers is effectively associated with a 
rearrangement of the amphiphilic molecules in the edge, as Glaser suggested.
Recently, Talanquer et al.~\cite{Talanquer_JCP_03} confirmed Eq.~(\ref{energy_pore}) 
using a density functional theory for amphiphilic bilayers.

Here we present a large scale study of spontaneously formed pores in a stack 
of tensionless bilayers. In stretched bilayers ($\gamma \neq 0$)
the pore energy mainly depends on the pore area, but  at zero surface tension ($\gamma = 0$),
the relevant variable is the contour length $c$ [see Eq.~(\ref{energy_pore})]. 
In particular, we study systematically the shapes of pores,  
their spatial distribution within a bilayer, and their dynamical evolution.

The paper is organized as follows: in Sec.~\ref{model_method} we describe 
briefly the model of amphiphilic molecules and the 
molecular dynamics simulation methods which permit us to
describe the lamellar phase in the $NPT\gamma=0$ ensemble 
(for more details see Refs.~~\onlinecite{Loison_JCP_03,Loison_thesis}). 
Then, the analysis of the molecular dynamics configurations is described. 
The results presented in Sec.~\ref{results} are divided into four parts.
First,  in Sec.~\ref{composition}, we present the density profiles around the pores. 
These show that we study hydrophilic pores, \textit{i.e.}~pores with a configurational 
rearrangement in the edge.
In Sec.~\ref{positions}, the pore positions within a membrane are investigated.
We conclude that the pores do not interact unless close to each other. 
In Sec.~\ref{size_shape}, we discuss the size  and shape distribution of the pores and 
estimate the line tension $\lambda$ associated to the configurational rearrangement in 
the pore edge. 
Finally, we trace the time evolution of individual pores. 
The results are described in Sec.~\ref{time_evol},
and interpreted with a simple stochastic model~\cite{Khanta_Pramana_83}. 
We summarize and conclude in Sec.~\ref{conclusions}.

\section{Model and Methods}
\label{model_method}

At the time-scale available to all-atoms molecular simulations (about $100\,$ns), 
the spontaneous formation of a pore in a  lipid bilayer can be considered as a rare event. 
In fact, the formation of a pore  in a DPPC bilayer was studied a few years ago 
with all-atom molecular dynamics \cite{Marrink_JACS_01} of 256 lipids during $0.5\,\mu$s,
but this pore was observed in a far-from-equilibrium situation with a relatively small system.
To explore the properties of thermally activated pores in equilibrium bilayers,
we use a coarse-grained molecular model which describes relatively thin bilayers. 
In such simulations, many pores appear spontaneously during one simulation run.

\subsection{Coarse-Grained Molecular Model}
The model was presented in detail in a previous publication \cite{Loison_JCP_03}
 and is based on well known similar models \cite{Kremer_JCP_90,Soddemann_EPJE_01}, therefore 
we recall here only its essential features. The bilayers are formed by amphiphilic molecules 
(molecules with a  hydrophilic head-group and one hydrophobic tails) in a binary 
solution with solvent.
All molecules are represented by one or  several soft beads (for simplicity, all 
beads are taken to have the same mass $m$).
The solvent is represented by single soft spheres (type $s$). 
 The amphiphilic molecules are linear tetramers composed of
 two solvophobic beads (or ``tail beads", denoted $t$) and
 two solvophilic beads (or ``head beads", denoted $h$).  
The soft spheres of the amphiphilic $h_2t_2$ are connected by bonds.

Bonded beads are connected by a spring potential \cite{Kremer_JCP_90} independent of the bead-types:
\begin{eqnarray}
\lefteqn{ U_{LJ-FENE}(r) =}\\ \nonumber \\
&\left\{
  \begin{array}{ll}
        4  \epsilon \left[\left(\frac{\sigma}{r}\right)^{12} 
    - \left(\frac{\sigma}{r}\right)^6\right] 
    - \left( \frac{\kappa  r_b^2}{2\sigma^2}\right)
    \ln\left[ 1-\left(\frac{r}{r_b}\right)^2\right]& 
    \mbox{if } r  \leq  r_b \\ 
    \infty & \mbox{if } r_b  \leq r \\
     \end{array}
   \right.&\nonumber 
\label{FENE_pot}
\end{eqnarray}
This potential comprises a Lennard-Jones type soft repulsive part dominating  for $r  \to 0$,
and an attractive part diverging  for $r  \to r_b$ (the bond length is therefore confined between 0 and $r_b$).
The length parameter $\sigma$ is our unit of length and
 the energetic  parameter $\epsilon$ our unit of energy.
The bond parameters were fixed at $r_b = 2.0 \, \sigma$ and 
$\kappa = 7.0 \, \epsilon $. 

Non-bonded beads interact with short ranged potentials of the general form 
 \begin{eqnarray}
\lefteqn{ U_{LJ-cos}(r)=}\\
 \nonumber \\
&\left\{
  \begin{array}{ll}
       4 \epsilon \left[\left(\frac{\sigma}{r}\right)^{12} 
       - \left(\frac{\sigma}{r}\right)^6 +\frac{1}{4}\right] 
       - \phi & \mbox{if } r  \leq 2^{1/6} \sigma\\ 
       \frac{\phi }{2}\left[ \cos(\alpha r^2+\beta) -1\right] 
       & \mbox{if } 2^{1/6} \sigma \leq  r\leq r_c \\
       0 & \mbox{if } r_c \leq r
     \end{array}
   \right. &\nonumber 
\label{LJ-cos}
\end{eqnarray}
This potential comprises  again a Lennard-Jones type soft repulsive part,
and a short-ranged attractive part. The range $r_c$ of the potential is fixed at $1.5\, \sigma$.
The parameters $\alpha$ and $\beta$ 
are fixed such that potentials and forces are continuous everywhere
(\mbox{$\alpha = \pi/r_c^2-2^{1/3} \sigma^2$} and \mbox{$\beta=2 \pi -r_c^2 \alpha$}).
The energetic parameter $\phi$ determines the depth of the potential; it depends on the 
types of the interacting beads. For pairs that ``dislike'' each other ($ts$ and $th$),  we choose
$\phi =0$ and so that the  interaction is purely repulsive. For the sake of simplicity, for all the other pairs
($ss, sh, hh$, and $tt$), the potential depth $\phi$ of the
pair interactions is the same ($\phi = 1.1 \,\epsilon$). For a fixed $\epsilon$, the self-assembly 
is driven  by the increase of  $\phi$ only. The choice of $\phi = 1.1 \,\epsilon$ ensures that 
we simulate the liquid crystalline lamellar phase of the tetrameric amphiphiles \cite{Loison_JCP_03}.

In the following, lengths shall be given in units of $\sigma$, 
energies in units of  $\epsilon$ and masses in  units of $m$. 
This gives the time unit $\tau =  (m \sigma^2 /\epsilon)^{1/2}$. 
 Typical orders of magnitude of our units are for example
$\epsilon \sim  5\cdot 10^{-21}\,$J, $m \sim 10^{-25}$ kg, $\sigma \sim 5 \AA$, and 
$\tau \sim 10^{-12}\,s$.   In this case, one solvent bead represents 
roughly three water molecules, and the tetrameric amphiphiles typically represent
 a symmetric neutral surfactant like  $C_{6}EO_3$. 
Such surfactants are much smaller than usual biological lipids.
As a consequence,  bilayers diluted in a large amount of solvent  are not stable in our model
(they separate into micelles). Thus, a quantitative comparison of our system with  
realistic  lipid biomembranes is not our goal: we focus on
the general properties of pores in bilayer membranes. 

\begin{figure}[h!]
\begin{center}
\includegraphics[width=6cm]{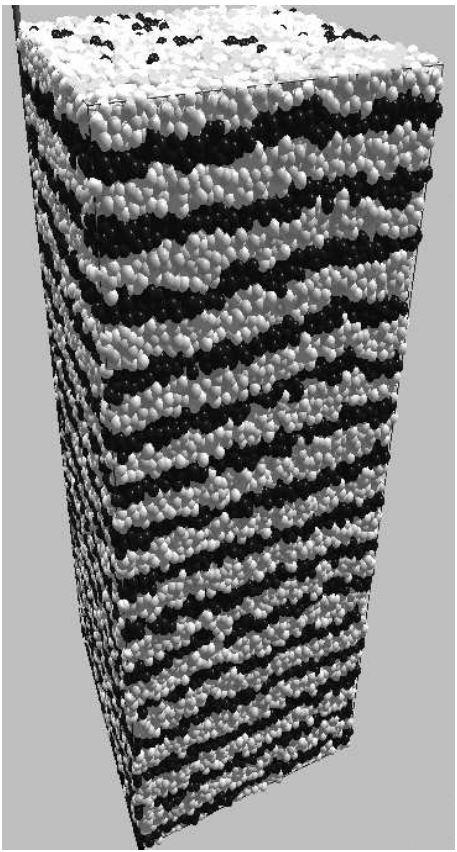}
\caption{Snapshot of the system (30\,720 $h_2t_2$ and 30\,720 solvent beads simulated 
in the NPT ensemble with $P/\epsilon =2.9~ \sigma^{-3}$, $k_BT = \epsilon$, and 
$\phi/\epsilon=1.1$). The dark beads are solvophobic (type t), 
the light beads are solvophilic beads or solvent beads (type h or s).  }
\label{lamellar_snapshot}
\end{center}
\end{figure}

To study stable bilayer membranes, we simulated self-assembled $L_\alpha$ lamellar structures:
a stack of amphiphile bilayers parallel to each other, separated by layers of solvent 
(see Fig.~\ref{lamellar_snapshot}).  
At the temperature of the simulation, these bilayers are two dimensional 
fluids~\cite{Loison_thesis, Loison_JCP_03}.
In the present work, a smectic phase composed of fifteen 
bilayers containing several thousands of molecules each was simulated over 
$10^5 \,\tau$. 

\subsection{Simulation Details}
We have analyzed the same simulation runs as in a previous related article~\cite{Loison_JCP_03}, 
in which the lamellar $L_\alpha$ phase  and its elastic properties are characterized in 
more detail. Here, we focus the analysis on the defects appearing in the bilayers due to 
thermal fluctuations.  We have studied the model in the $(N P_n P_t T)$-ensemble (constant 
number of particles, constant pressure normal and tangential to the bilayers, and constant 
temperature) with molecular  dynamics simulations. 
($N$): The lamellar phase was studied at an amphiphile fraction of $80 \%$ of 
the beads (one solvent bead per $h_2t_2$). The system contained $30\,720$ tetramers and 
$30\,720$ solvent beads, which formed fifteen parallel bilayers of about two thousand molecules 
each (see Fig.~\ref{lamellar_snapshot}).  

($P$): The normal and tangential pressure components $P_n$ and $P_t$ were
 kept constant using the extended Hamiltonian method 
of Andersen \cite{Andersen_JCP_80,Parrinello_PRL_80}.
The box  shape is constrained to remain a rectangular parallelepiped. 
The box dimension perpendicular to the bilayer ($L_z$) 
and tangential to the bilayers ($L_x, L_y$) are coupled to two separated pistons.
The thermal-averaged box dimensions were $ L_x =  L_y = 43.4 \pm 0.1 \,\sigma$ 
and $L_z =  31.9 \pm 0.1 \,\sigma$.
We imposed the two pressure components separately rather
than the total pressure $P = (P_n+2P_t)/3$ because of technical reasons: 
the mechanical equilibrium  is reached earlier,
the orientation of the bilayers is stabilized, and the surface tension is controlled.
Since we studied a bulk  lamellar phase, 
we imposed an isotropic pressure  ($P_n = P_t= P$). 
More details on the simulation algorithm and its parameters are given in 
Ref.~~\onlinecite{Loison_JCP_03}.  
($T$): The temperature was controlled by a stochastic
Langevin thermostat that  has been described earlier and applied
to very similar models\cite{Kremer_JCP_90,Soddemann_EPJE_01, Kolb_JCP_99}.

The dimensionless pressure and temperature were fixed  at $ P \sigma^3 /\epsilon = 2.9$ and 
$k_BT/\epsilon = 1.0$. For the chosen dimensionless potential depth, $\phi/\epsilon= 1.1$, 
the density fluctuates around $0.85$ beads per unit volume. 
With these typical parameters, lamellae form and order spontaneously. 
But this process  requires at least $30\,000\,\tau$.
We have  therefore imposed the orientation of the lamellae in the initial configurations. 
They were constructed so, that fifteen bilayers separated by solvent 
layers were stacked in the $z$-direction. These configurations were then 
relaxed for $10\,000\,\tau$. During that time, the interlamellar distance adjusted 
to its equilibrium value, the shape of the flexible box changed accordingly,
but the director remained basically aligned with the $z$-direction.
 
Data were then collected over $10^5\,\tau$. We verified that the pressure tensor 
obtained at equilibrium was diagonal and isotropic, 
and  that the surface tension $\gamma =\langle L_z (P_n-P_t)\rangle$ 
was negligible \cite{Loison_JCP_03}. For the time independent analysis (Secs. \ref{composition}, \ref{positions} and 
\ref{size_shape}), we have used  400 configurations separated by $250\,\tau$. 
For the time dependent analysis (Sec. \ref{time_evol}), we have used four series 
of 400 configurations separated by $1\,\tau$. These series were separated by $10\,000\,\tau$.

\subsection{Pore Detection and Analysis}
\label{pore_detection}
Generally, several types of point defects exist  in smectic phases. 
A highly aligned $L_\alpha$ phases may contain pores in the bilayers, but also necks or
passages between two bilayers \cite{Bouligand_LC_99,Constantin_PRL_00}.
Any of those defects changes the topology of the lamellar phase:
 pores in the bilayers connect two neighboring solvent layers, 
fusion between neighboring bilayers connect the bilayer matrices, and 
passages connect both neighboring solvent layers and neighboring bilayers. 
One possibility to detect a defect, and to  distinguish between these defects is to analyze the 
topology of the lamellar phase \cite{Constantin_PRL_00,Holyst_JCP_97}. Using a cluster-algorithm on the 
simulated lamellar phase \cite{Loison_thesis}, we observed that 
in our model $L_\alpha$ phase, necks and passages
appear very rarely (in less than 1\% of the configurations). 
So we can focus on the pores in the bilayers. 

\begin{figure}[h!]
\begin{center}
\includegraphics[width=6cm,height =6cm]{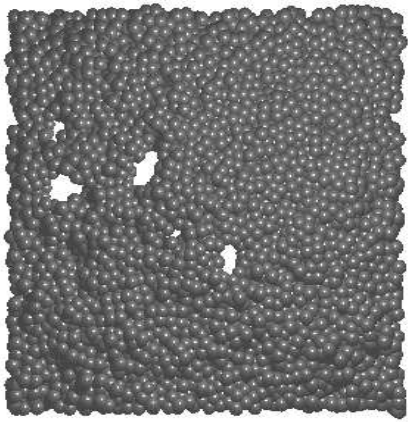}
\caption{Snapshot of the head-beads of one bilayer (top view).}
\label{pore_snapshot}
\end{center}
\end{figure}

Figure \ref{pore_snapshot} shows a representative snapshot of the tail-beads of one membrane.
On the snapshot, five pores are clearly present.

For a systematic analysis, we defined a pore as a region of the inner part of the
bilayer were relatively few tail-beads are present (\textit{i.e.} the relative density of tail beads is
less than a given threshold).  The $n$th bilayer is defined by its position 
$h_n(x,y)$  and its thickness $t_n(x,y)$ (Monge representation \cite{Safran_94}).
In practice, only discrete values of $x$ and $y$ were considered 
($x = n_x {L_x/N_x}$ and $y = n_y {L_y/N_y}$ where $n_{x,y}$ are integer values, $N_x=N_y=32$,  and 
$L_{x}=L_{y} \sim 43\,\sigma$). Therefore, all the observables are measured on a two dimensional 
grid of $32 \times 32$ ``plaquettes" of area $d_x\times d_y$ where $d_x=L_x/N_x$ and $d_y=L_y/N_y$.
 As the dimensions of the simulation box fluctuate, the mesh size also fluctuates:
 $dx = dy \sim 1.3  \pm 0.05\,\sigma$.
In the following, the notation  ``plaquette" indicates that
area are expressed in units of the plaquette area and the lengths in units of the mesh size. 

The analysis is divided into two steps 
(see Appendix  \ref{appendix_pores} for more details):
First, we have determined the local positions and thickness of the membranes in every configuration from 
the relative densities of solvophobic beads $\rho_{tail}(x,y,z)$.
For each point  $(x,y)$, the relative density of tail beads $\rho_{tail}(x,y,z)$ oscillates as a function of $z$, with
one maximum per bilayers. The position $h_n(x,y)$  and thickness $t_n(x,y)$ of a given membrane were determined 
as the mean and the difference of the two $z$ values where $\rho_{tail}(x,y,z)$
 equals a threshold value  ($\rho_0\sim 0.7$) around the appropriate maximum. 

Second,   the positions $(x,y)$  where the thickness  of the membrane $n$ is zero or undefined 
are attributed to pores. We call them ``pore-positions".
For each membrane, the pore-positions $\{ x_i,y_i\}^{pp}_n$ 
are assembled into two dimensional clusters, the ``pore-clusters".
Two pore-positions belong to the same pore-cluster if they share at least one vortex of 
the two-dimensional  $N_x\times N_y$ grid. 

\begin{figure}[h!]
\begin{center}
\includegraphics[width=6cm,height =6cm]{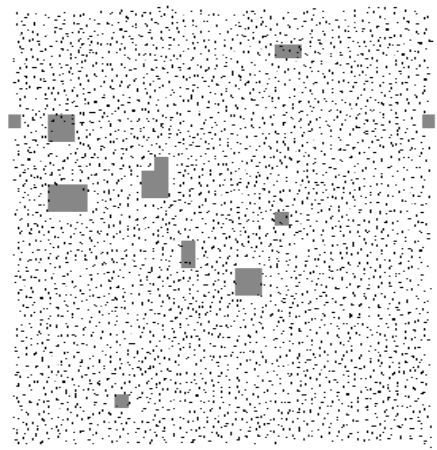}
\caption{Result of the pore analysis on the membrane depicted in Fig.~\ref{pore_snapshot}.
The single isolated square in the bottom left of the image is of the area of a single plaquette. }
\label{pore_analysis}
\end{center}
\end{figure}

The result of such an analysis is represented on Fig.~\ref{pore_analysis}.
In this figure, the dots represent the centers of the tail-beads 
belonging to the membrane (seen in Fig.~\ref{pore_snapshot}).
Each pore-position is represented by a grey square. The resulting grey patches are the pore-clusters. 
Each pore visible in Fig.~\ref{pore_snapshot}  appears under the form of a pore-cluster.
But additional small pores-clusters that were not visible on the snapshot also appear in the analysis.
It seems that some tail-beads (black dots) are still present in those pores.
In practice, it was difficult to decide whether these small pore-clusters are noise, fluctuations of the bilayer 
thickness, or  hydrophobic pores.  In particular, no minimum life-time was found for these pores of minimum size.
Therefore,  the pore-clusters composed of one plaquette only were disregarded
 in the following analysis (unless noted otherwise).

For each of the pore-clusters, which we call only ``pores" for simplicity, 
the mean position ${\mathbf r}_{cm}$, the matrix of gyration $g^{\alpha \beta}$, 
the area $a$ and  the  contour length $c$ were calculated.
 Again, these observables are defined for clusters of pixels on the two dimensional  
$N_x\times N_y$ grid. The mean position is calculated via
\begin{equation}
{\mathbf r}_{cm}=\frac{1}{n_{pp}}\sum_{i=1}^{n_{pp}} {\mathbf r}_i,
\end{equation}
where the sum runs over the $n_{pp}$  pore-positions  of the pore cluster, 
and ${\mathbf r}_i$ is the position of the pore-position number $i$. 

The gyration matrix is given by
\begin{equation}
\label{gyration}
g^{\alpha \beta} = g_0 + \frac{1}{n_{pp}}\sum_{i=1}^{n_{pp}} ({\mathbf r}_i - {\mathbf r}_{cm})^\alpha ({\mathbf r}_i - {\mathbf r}_{cm})^\beta,
\end{equation}
where $\alpha$ and $\beta$ denotes axes of the grid, and $g_0$ the gyration matrix 
of one plaquette, which we approximated by one fourth the identity matrix (in plaquette area).

\section{Simulation Results}
\label{results}

In each bilayer of area $L_xL_y \sim 1850\,\sigma^2$, the average number of pores is 
$9.9 \pm 1.3$. Their total area 
corresponds to approximatively $1\%$ of the projected area of the bilayer. 
Among these  ten pores, $5.7 \pm 1.0$ pores on average have the minimum size of one 
plaquette (with an area of about $1.7\,\sigma^2$).

\subsection{Composition Profiles through the Pores}
\label{composition}
\begin{figure}[h!]
\begin{center}
\includegraphics[width=6cm,angle =-90]{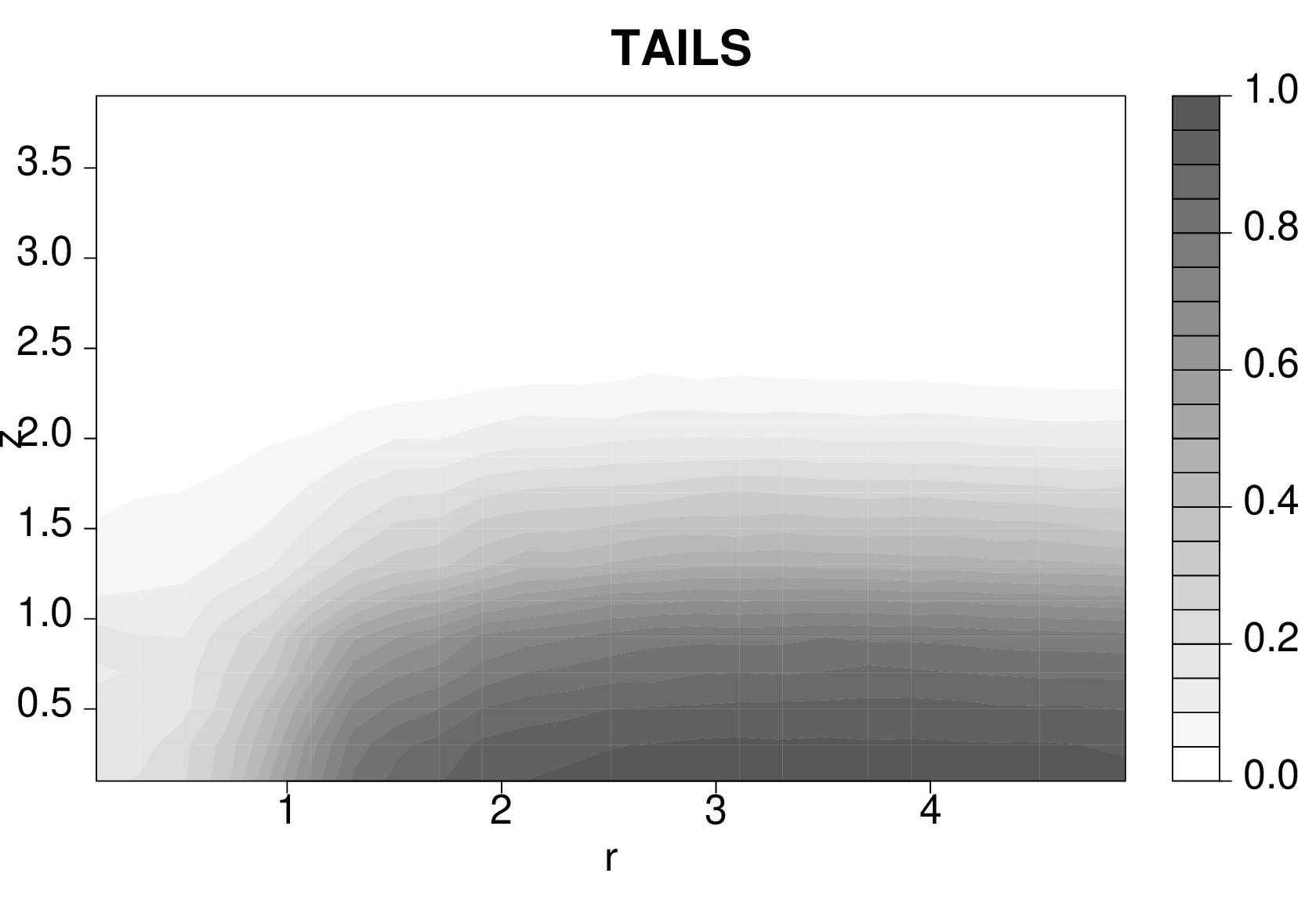}
\includegraphics[width=6cm,angle =-90]{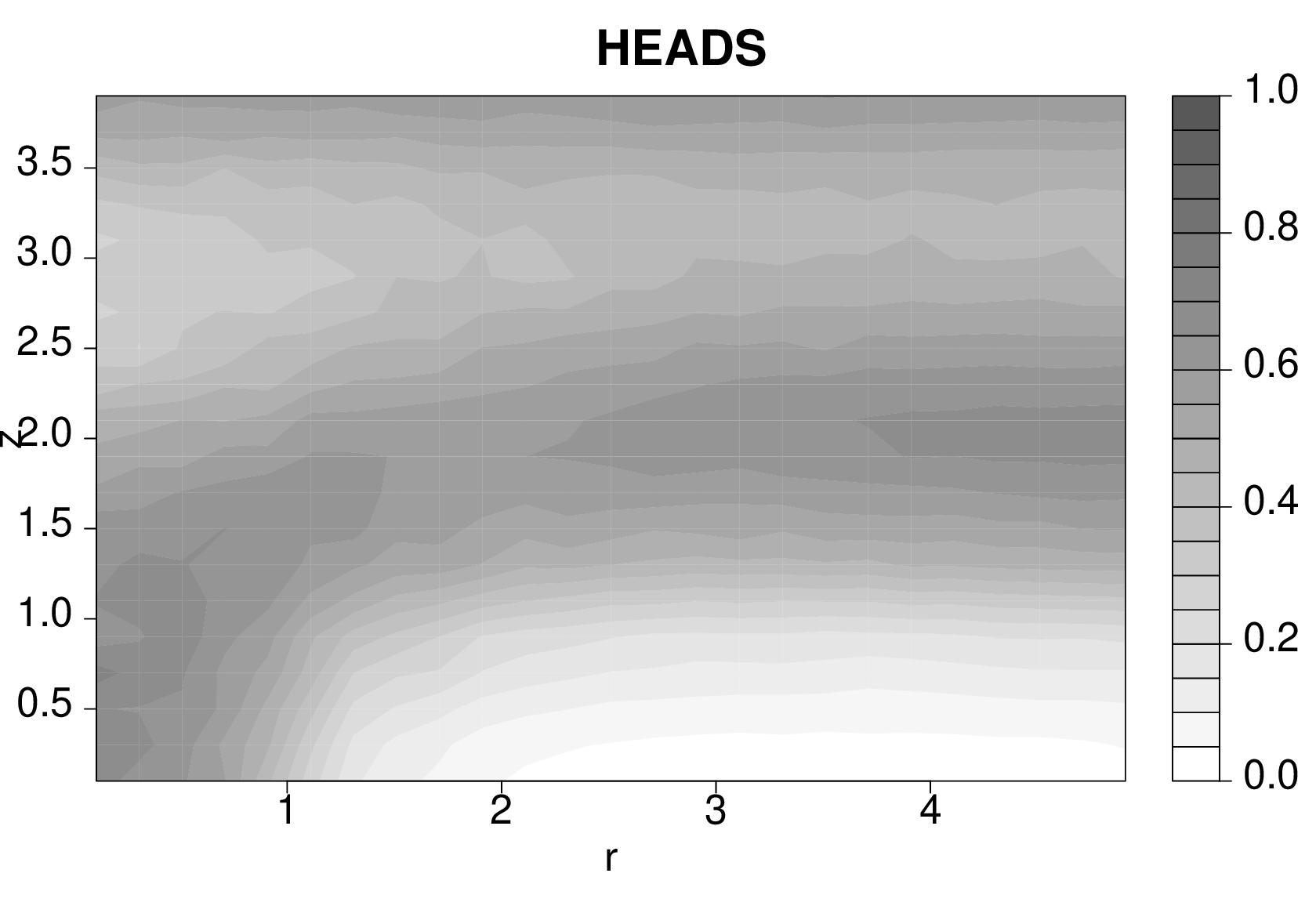}
\includegraphics[width=6cm,angle =-90]{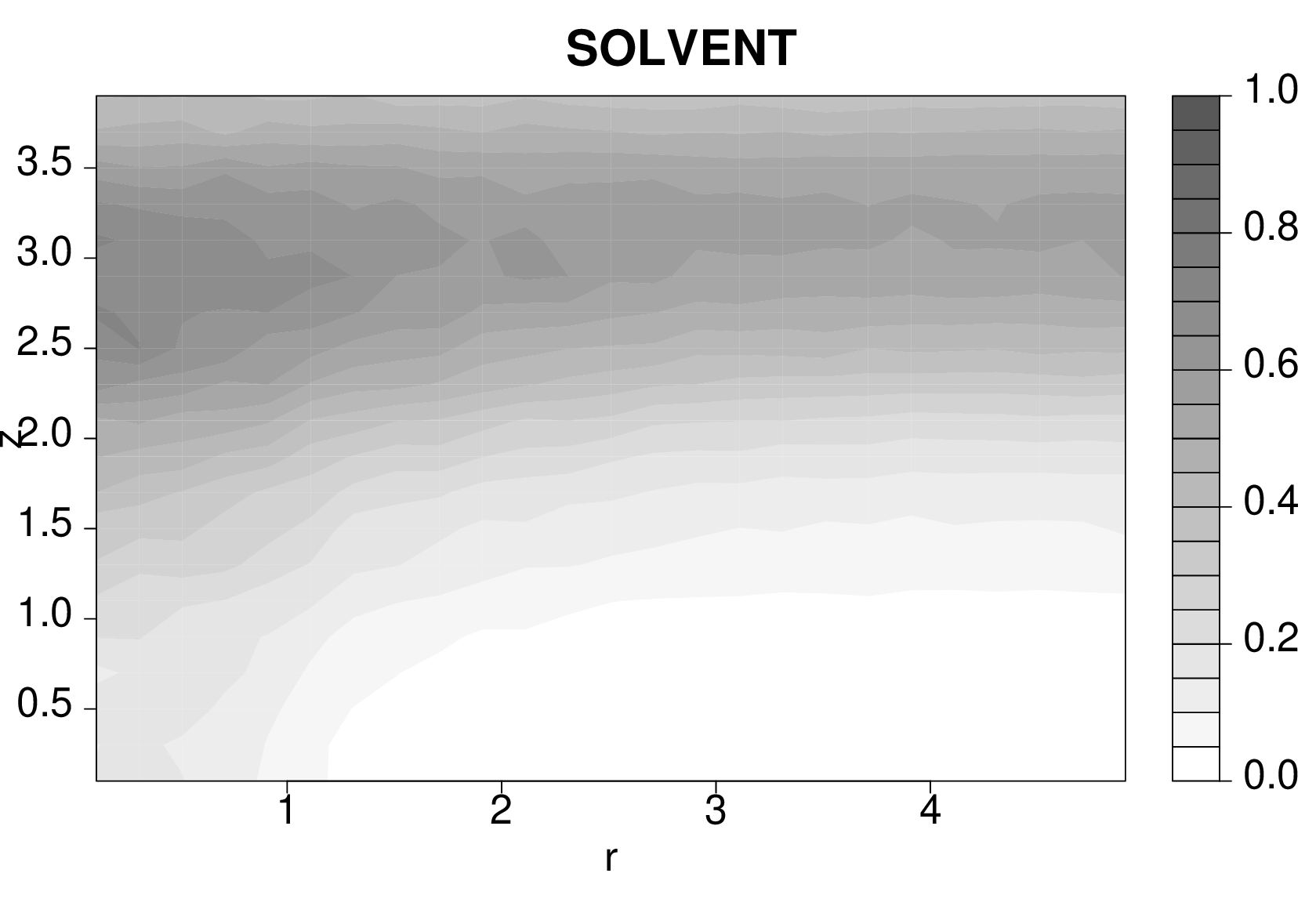}
\caption{Composition profiles around the centers of pores 
with areas between two and four plaquettes (length unit: $\sigma$). 
Note that the inter-bilayer distance is $6.4\,\sigma$,  
so the mid-bilayer plane (between a  bilayer and its next neighbor) is at $z=3.2\,\sigma$}
\label{density_profiles}
\end{center}
\end{figure}

Figure \ref{density_profiles} shows the composition in head, tail and solvent beads around the 
center of relatively small pores  (of area two to four plaquettes). They 
are represented in  the cylindrical coordinates with the center of the pore as origin:
 $\rho_{tail}({\mathbf r},z)$, $\rho_{head}({\mathbf r},z)$, $\rho_{solvent}({\mathbf r},z)$,  
where the $z$-axis is perpendicular to the bilayers (see Fig.~\ref{hole_types}). 
We averaged over the directions of $\bf{r}$ in the plane of the bilayer and 
over the two sides of the bilayers for the values $+z$ and $-z$, 
so the composition is represented as a function of $|r|$ and $|z|$.

Far from the pore centers, at distances $r\geq4\,\sigma$, the typical bilayer structure 
is found.  The tails are segregated inside the bilayer ($z \leq 2\,\sigma$). 
For intermediate z (around $2\sigma$), the head-beads shield the tail-beads from the solvent.  
The solvent is concentrated near the mid-bilayer plane ($z =3.2\,\sigma$).
As defined by our detection algorithm, the pore  ($r \to 0, z \leq 3\sigma$) is 
characterized by the absence of tail-beads inside the bilayer. 
The head and solvent distributions show that the interior of the pore is occupied 
by head-beads rather than solvent-beads. Even inside the pore, the head-beads shield 
the tail-beads from the solvent. We conclude that pores with an area $a$ larger than 
two plaquettes are of ``hydrophilic type". As expected, the same observation is made for 
larger pores (data not shown). 
To summarize, the amphiphiles of the pore edge are reoriented in our model, even for 
the small pores ($a=2,3,4$ plaquettes). Notably, such a  reorientation was also observed
by M\"uller and co-workers in the pore edges of amphiphilic bilayer of 
block-copolymers~\cite{Mueller_JCP_96}.
As a consequence, it is reasonable to use Eq.~(\ref{energy_pore}) to interpret our 
simulation data: the energy of a pore is expected to increase with the contour 
length of the pore.

\subsection{Positions Correlations of the Pores}
\label{positions}

About four pores of area larger or equal to two plaquettes are present in each simulated  
membrane of area $L_x\times L_y \sim 43 \times 43 \,\sigma^2$. Do these pores interact? 
A priori, they are at least subject to hard core repulsion, because two pores
cannot overlap (overlapping pores would be considered as a single pore). In this section, we try to 
detect whether there exist additional soft interactions
between the pores.

\begin{figure}[h!]
\begin{center}
\includegraphics[width=6cm,angle =-90]{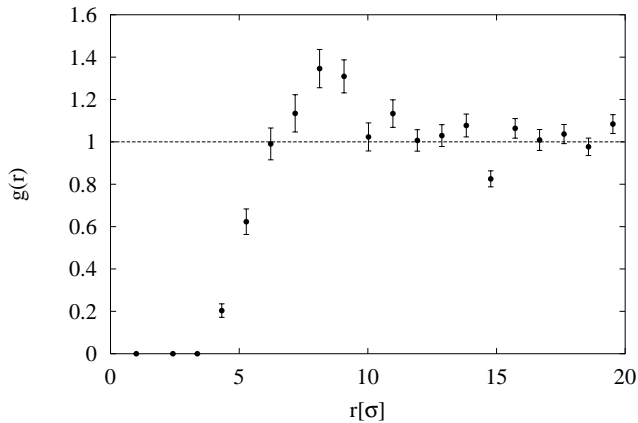}
\caption{Spatial pair correlation function of the center of the pores. 
As we neglected the correlations between the distributions
of pores in different membranes of the same configuration, the errorbars are under-estimated.}
\label{gr}
\end{center}
\end{figure}
The most straightforward approach to this problem is to look for spatial pair correlations 
between the pores (see Fig.~\ref{gr}).
In this analysis, we have taken into account all the pores, even the smallest ones 
($a=1$ plaquette). Despite the noise of the data, two tendencies are clear. 
At large distances, say  about $r \geq 10 \,\sigma$, no correlation is observed.
At small distances  ($r \leq 5 \,\sigma$), the short-range repulsion between the 
pores is reflected by a depletion in the pair correlation function.  
At intermediate distances ($r \sim 8 \,\sigma$), the data suggest a slight positive
correlation, \textit{i.e.} the probability that pores open up seems to be slightly increased 
in the vicinity of other pores. The statistics is however not sufficient to justify a conclusive statement.

\begin{figure}[h!]
\begin{center}
\includegraphics[width=3cm,angle=-90]{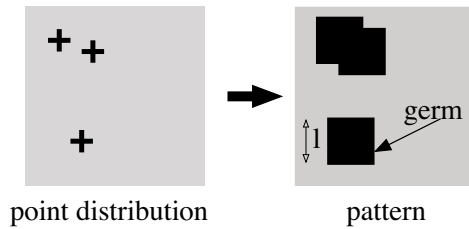}
\caption{Principle of the Minkowski analysis (see text for more explanations).}
\label{minkowski_principle}
\end{center}
\end{figure}

An alternative, complementary method of analyzing spatial distributions of patterns
in physical systems has been proposed by Mecke and 
others~\cite{Mecke_IJMPB_98,Brodatzki_CPC_02,Mecke_Lectures_00,Michielsen_PR_00,Mecke_Lectures_00_2,Mecke_lectures_02}: the evaluation of Minkowski functionals.
Minkowski functionals have been used in various areas of mathematics, chemistry 
and physics  to analyze high-order correlations of spatial distributions.
The present  analysis is similar to the one developed and  applied by Mecke, Jacobs 
and co-workers \cite{Mecke_IJMPB_98,Jacobs_langmuir_98} to study the spatial
distribution of defects in a thin film of polymers. The analysis can be  
decomposed into two steps (see Fig.~\ref{minkowski_principle}).
First, on each of the $N$ points of the distribution, a square of size $l$ is fixed 
(the germ)~\cite{footnote1}.
These $N$ germs form a two dimensional image, the pattern.
In two dimensions, the three dimensionless Minkowski functionals  $m_{0,1,2}$
of a pattern are proportional to its surface $A$, contour length $C$ and  
Euler characteristic $\chi$\cite{Mecke_Lectures_00}:
$m_0 = A/A_t$ where $A_t$ the total area of the surface, 
$m_1 = C/\sqrt{A_tN}$, where $N$ is the number of points in the distribution, 
and $m_2 = \chi/N$.
We used an algorithm described by Milchielsen \cite{Michielsen_PR_00} 
to calculate the Minkowski functionals of a digitalized pattern 
(without periodic boundary conditions). 
Finally, the  dimensionless Minkowski functionals are represented  as a function 
of the dimensionless  coverage factor $x = l^2 N/A_t$ where $l^2$ is the area of 
the square germ.

\begin{figure}[h!]
\begin{center}
\includegraphics[width=6cm,angle=-90]{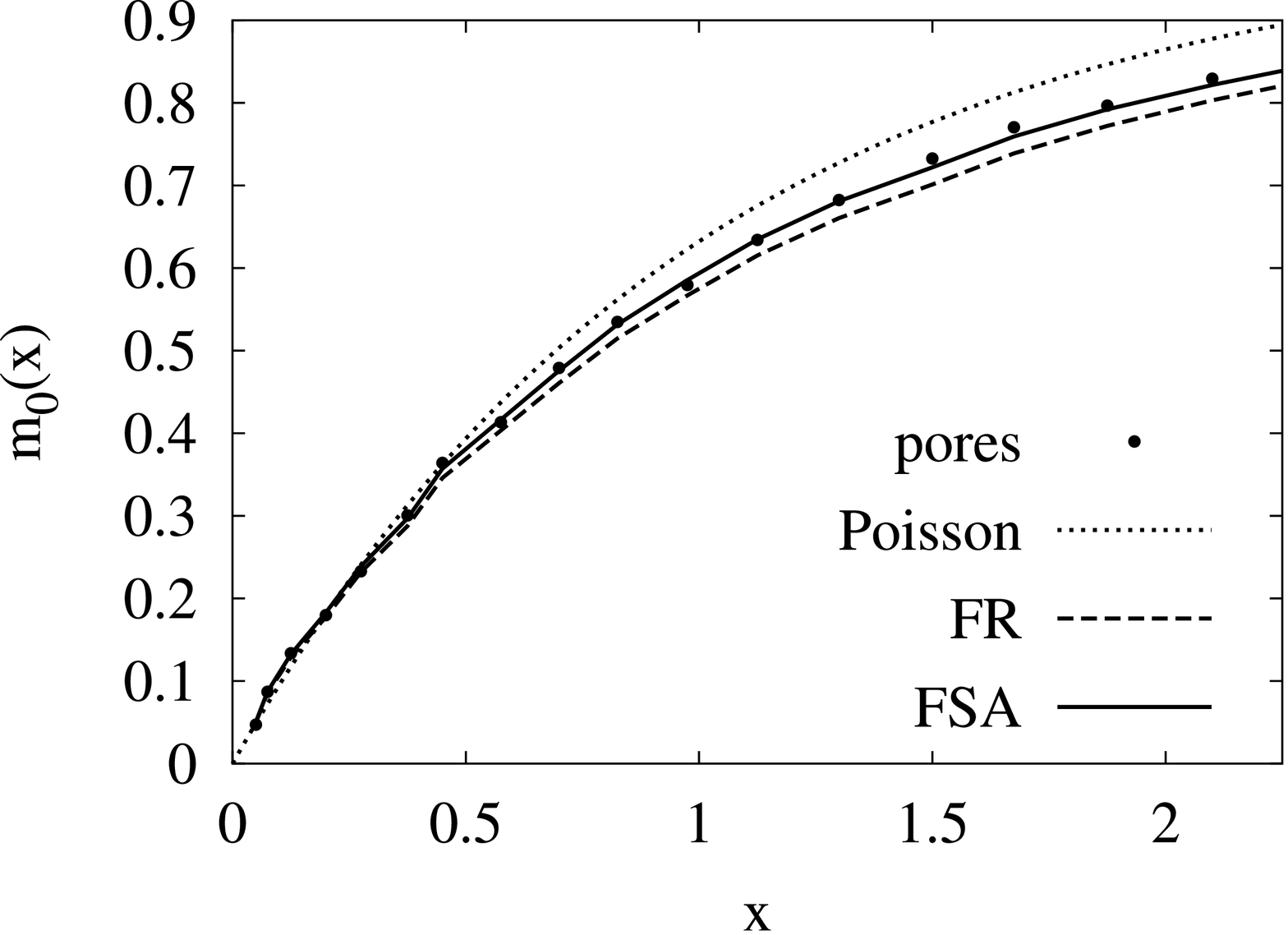}
\includegraphics[width=6cm,angle =-90]{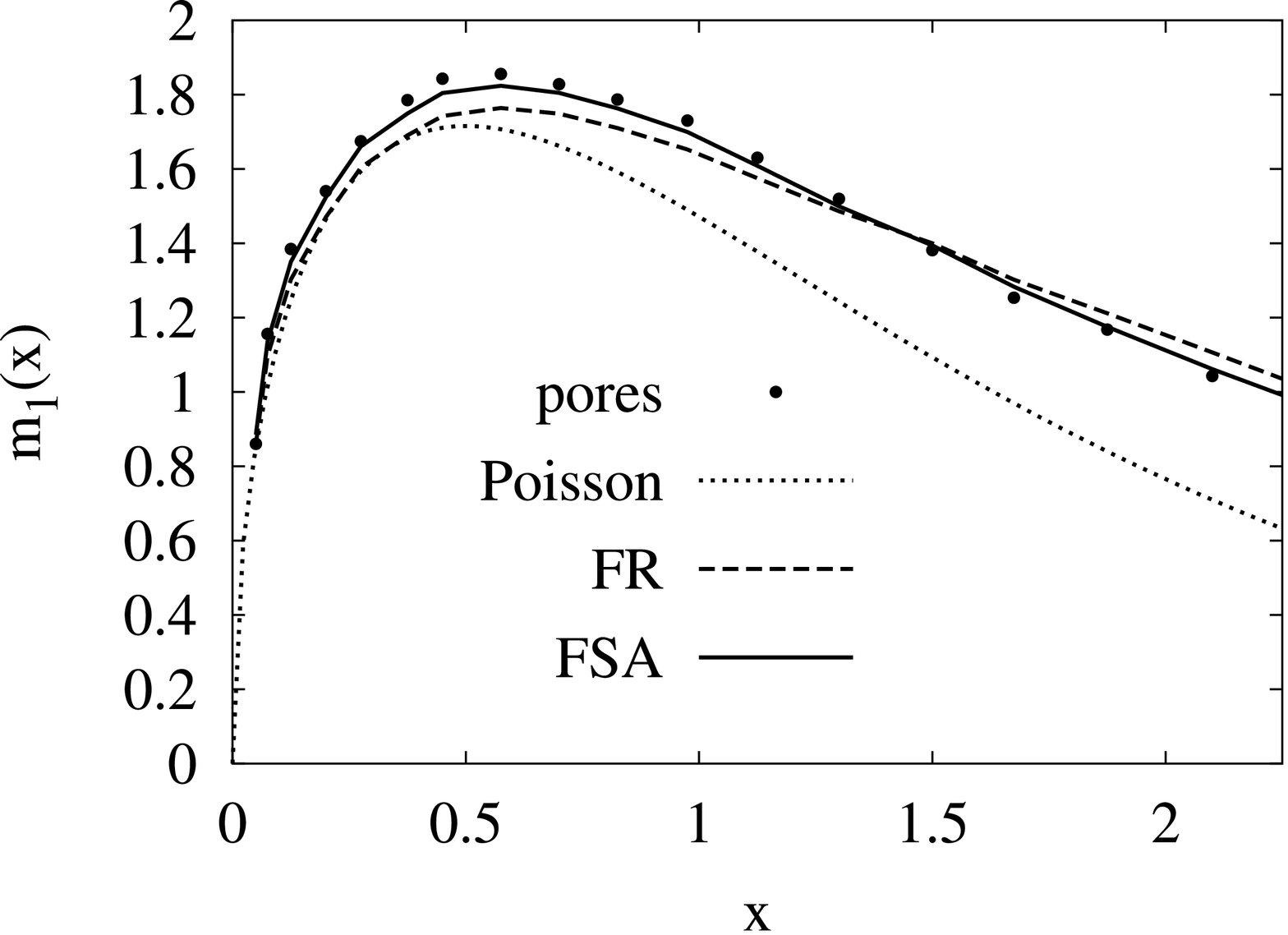}
\includegraphics[width=6cm,angle=-90]{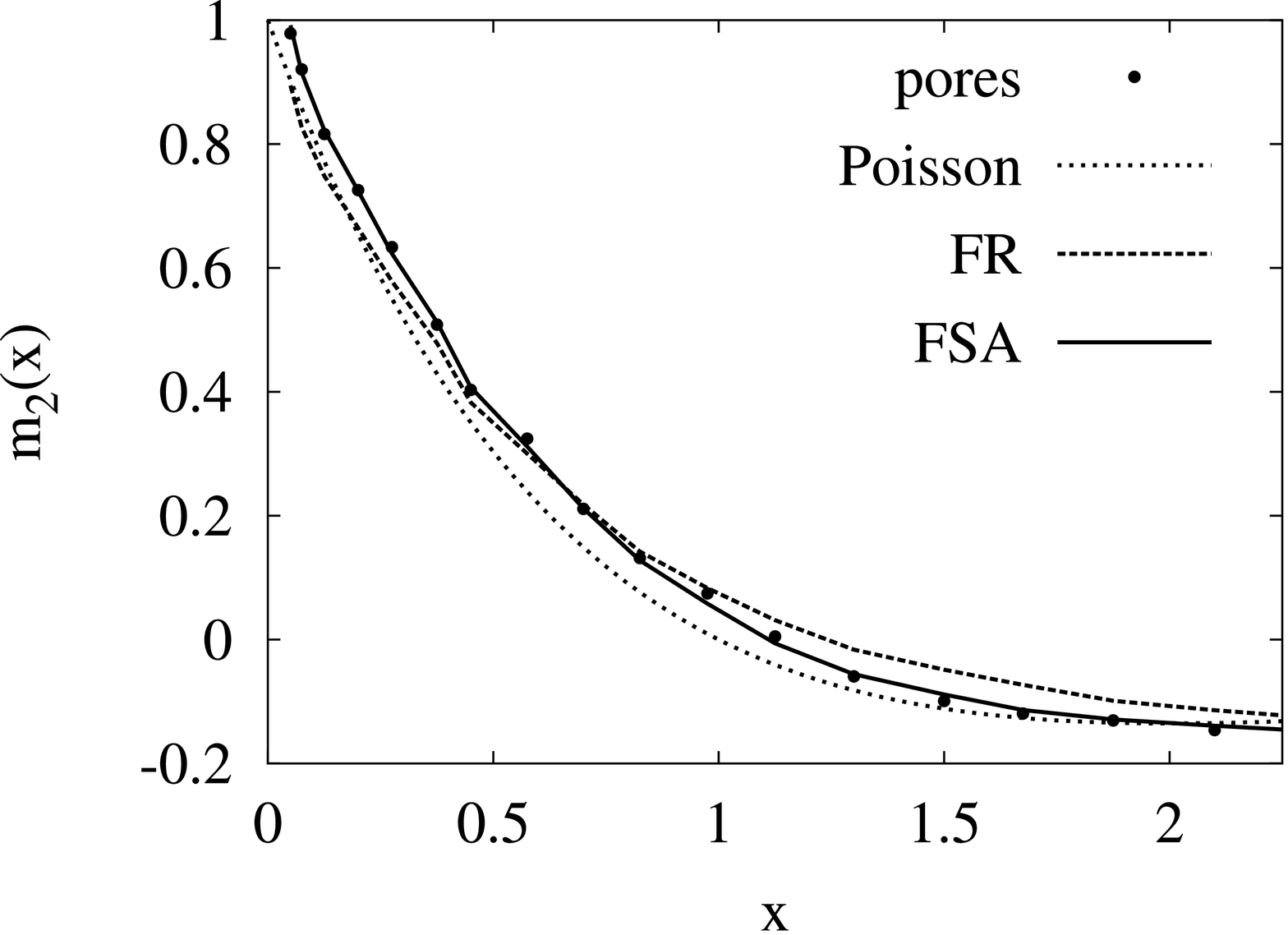}
\caption{Dimensionless Minkowski functional as a function of the dimensionless 
coverage factor (6 points on a $32\times 32$ grid). ``Poisson" corresponds to Eq.~\ref{minkowski_poisson}, 
FR to the numerical results obtained with 1000 Finite Random distributions,
 and FSA to the numerical results obtained with 1000 Finite Self-Avoiding distributions (see text and Appendix).}
\label{minkowski}
\end{center}
\end{figure}

Figure \ref{minkowski} shows the three Minkowski functionals obtained from about 1000 pore 
distributions containing \emph{exactly} six pores of area larger than two plaquettes (symbols).
 We compare these results with three reference distributions (curves).

The first reference is a random distribution of fixed germs on an infinite surface (Poisson), 
for which the Minkowski functionals are known analytically \cite{Mecke_Lectures_00}:
\begin{subequations}
\label{minkowski_poisson}
\begin{eqnarray}
m_0(x)&=& 1-\exp(-x) \label{equationa}\\
m_1(x)&=&4\sqrt{x}\exp(-x) \label{equationb}\\
m_2(x)&=&(1-x)\exp(-x) \label{equationc}
\end{eqnarray}
\end{subequations}
The pore distributions obviously differ from a Poisson distribution on an infinite surface.
We attribute the discrepancy to two effects: finite size effects, and repulsion effects.
The former are due to the finite size of our quadratic grid, and to the
finite area of the membranes. Explicit expressions for the Minkowsky functionals in a finite 
``observation window'' on an infinite surface are given in Ref.~~\onlinecite{Mecke_IJMPB_98}.
Here, we have additional boundary effects because only germs inside the window contribute
to the pattern. Therefore, we have calculated the expected finite size effects numerically.

Two reference distributions  on a finite grid  of $32 \times 32$ pixels were considered.
In the first one, which we call ``finite random'' (FR) distribution, a fixed number of points was distributed
randomly. The second one, denoted ``finite self-avoiding'' 
(FSA) takes into account the hard-core repulsion between pores, \textit{i.e.} pores may
not touch or overlap (see Appendix).  As Figure \ref{minkowski} shows, the differences between FR and FSA 
distributions are relatively small, but observable. We present here the results for bilayers containing 
six pores, but the same conclusions are drawn for different densities (the data are not shown; 
as expected, the influence of self-avoidance increases with the density of points).

The results obtained with the pore 
distributions of the simulations are closer to those obtained with the FSA distributions.  
Therefore, both finite size effects and self-avoidance are observable in the Minkowski analysis, and
these two effects seems to be sufficient to account for the data.

To conclude, the pores are not distributed randomly in the membrane.
The spatial distribution of pores is compatible with a simple model, in which 
the pores interact only through a hard-core repulsion.
Additional soft interactions between pores may be present,
but these do not influence the overall pore distribution significantly.

We have also studied the correlations between pores in different membranes.
Not surprisingly, a small correlation is observed, \textit{i.e.} the probability
that a pore opens up on top of another pore is slightly increased. 
But we did not explore this effect throughout.

\subsection{Size and Shape of the Pores}
\label{size_shape}
In the following the area of the pores, their contour length and their radius of gyration 
are analyzed. In this section, all pores have been taken into account, 
even the smallest ones ($a=1$ plaquette).

As shown in Section \ref{positions}, the spatial distribution of the pore within a membrane
were in good agreement with the hypothesis that the pore interact only via a 
hard-core repulsion.
In the following, we consider the pores as independent. Within this hypothesis, 
the contour length distribution $P(c)$ of the pores  can be compared to the
Bolzmann factor $ g(c) \exp[- E(c)/(k_BT)]$,
where $E(c)$ is the energy to create a single pore of contour length $c$ and $g(c)$ 
is the degeneracy. We calculated $g(c)$ for the particular quadratic grid of our analysis 
(see Table \ref{gc_table}). 

\begin{figure}[h!]
\begin{center}
\includegraphics[width=5.5cm,angle=-90]{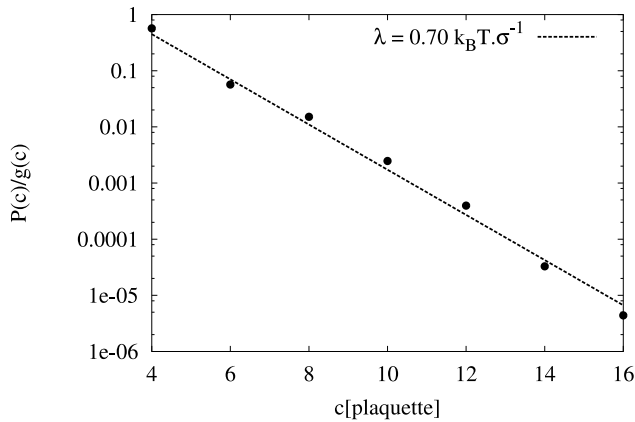}
\caption{Probability distribution function of the contour of the pores $P(c)$, 
divided by the degeneracy of the contour length $g(c)$ (linear-log plot).}
\label{N15_NX32_cg}
\end{center}
\end{figure}

The ratio $P(c)/g(c)$ is shown in  Fig.~\ref{N15_NX32_cg}  in a linear-log plot.
The approximated energy  of pore formation $- \ln\left[P(c)/g(c)\right]$ is  well described by a 
linear function.  Fitting the model $E(c) =E_0 +\lambda c$ to the data yields the  line tension 
$\lambda = 0.7 \pm 0.1 \,k_BT\cdot\sigma^{-1}$. This value is approximately 
$5\cdot10^{-12}\,J\cdot m^{-1}$, which agrees reasonably with the values calculated by 
May~\cite{May_EPJE_00} for the excess free energy of the packing rearrangement of amphiphiles 
in the edge.
Previous results \cite{Zhelev_BBA_93,Mueller_JCP_96,Moroz_BJ_97} report line tensions in the 
range of  \mbox{$10^{-11}\,J\cdot m^{-1}$ to $3\cdot10^{-11}\,J\cdot m^{-1}$}, \textit{ i.e.}
larger than the present value.
In these references, the line tension includes also  the  excess free energy necessary to
transfer the amphiphiles  from a reservoir to the edge of  pore
 (the chemical potential of the amphiphiles in the grey region of Fig.~\ref{hole_types}).
 This contribution (typically $10^{-11}\,J\cdot m^{-1}$) 
is proportional to  the surface tension of the bilayer \cite{Mueller_JCP_96,May_EPJE_00} 
and vanishes in the present case.
\begin{table}[h!]
\centering 
\begin{tabular}{l| r r r r r r r }
c   	& 4 & 6	& 8	& 10	& 12	&14		&16 \\
\hline
g(c) & ~~~~1 & ~~~~2	& ~~~~9	&~~~36	& ~~168	& ~~715	& ~2000\\
\end{tabular}
\caption{Degeneracy of the contour length $c$ of 
clusters of pixels connected by at least one vortex. 
As explained in Section \ref{pore_detection}, a pore is defined 
as a cluster of pixels connected by at least one vortex. In our calculations,
the directions $-x$, $x$, $-y$ and $y$ are distinguishable. 
We neglect the finite size of the grid (a square of $32 \time 32$ plaquettes), 
therefore all positions of the pores on the grid are considered as equivalent. 
 Because of the high cost of such calculations, the value are exact only
 up to $c=14$. The value for $c=16$ is an under-estimate.}
\label{gc_table}
\end{table}

In brief, the size distributions computed with the simulations are compatible with
the usual mesoscopic model of pores energetics and
permit to compute the approximate line tension of the pore edge.

\begin{figure}[h!]
\begin{center}
\end{center}
\includegraphics[width=5.5cm,angle=-90]{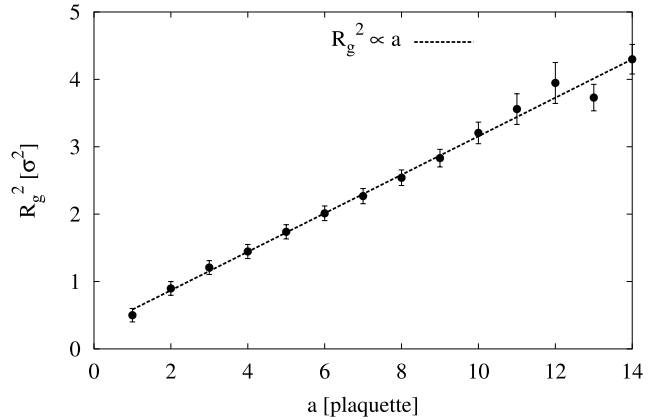}
\caption{ Square of the  radius of gyration of the pores, as a function of their area.}
\label{N15_NX32_Rg_A}
\end{figure}

The shape of the pores have been studied with the two-dimensional gyration matrix
of the pore-clusters [see Eq.~(\ref{gyration})]. The two (positive) eigenvalues of the gyration 
matrix are denoted $\rho_1^2$ and $\rho_2^2$. The sum of the eigenvalues 
$R^2_g = \rho_1^2 +\rho_2^2$ is the square of the radius of gyration of the pore, 
and the relative difference $|\rho_1^2 -\rho_2^2|/R^2_g$,  its asymmetry
- it is zero for a circular pore and tends towards 
one when the pore is elongated in one direction.
Figure \ref{N15_NX32_Rg_A} represents the radius of gyration
 of the pores as a function of their area.
As expected, the square of  the radius of gyration increases
 linearly with the area of the pores. A least-square linear fit yields 
\mbox{$R_g^2 \sim 0.17 a + 0.29$ (both in units of $\sigma^2$)}.
The proportionality factor ($0.17$) is slightly  larger than 
the proportionality factor obtained for a homogeneous
disc (\mbox{$[2\pi ]^{-1} \sim 0.15$}).

\begin{figure}[h!]
\begin{center}
\end{center}
\includegraphics[width=5.5cm,angle=-90]{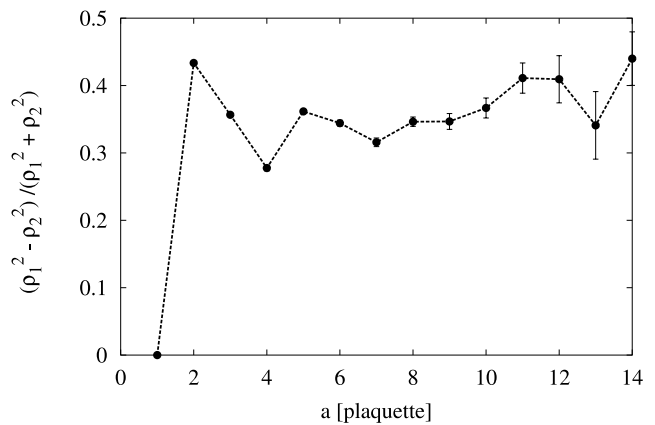}
\caption{ Asymmetry of the pores as function of their area. }
\label{N15_NX32_Rg_A_assym}
\end{figure}
As illustrated in Fig.~\ref{N15_NX32_Rg_A_assym}, 
the asymmetry of the pores does not vary significantly 
with the size of the pores, except for the smallest pores, whose asymmetry is
imposed by the finite mesh size of the grid.

This suggests that only one length-scale is sufficient to describe the pore dimension.
Notably,  the average asymmetry is not zero ($0.35 \pm 0.05$).
In fact, since the bilayers are flaccid, there is no reason why the pores should be circular. 
M\"uller and co-workers \cite{Muller_JCP_02} emphasized the importance of the asymmetric 
shape of pores on the mechanism of fusion between two parallel bilayer membranes.

\begin{figure}[h!]
\begin{center}
\end{center}
\includegraphics[width=5.5cm,angle=-90]{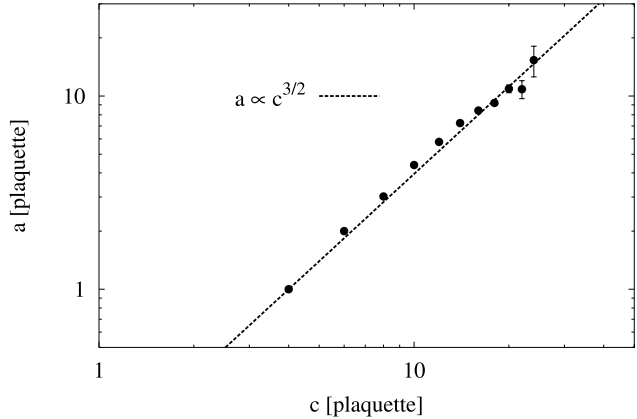}
\caption{Area of the pores as a function of their contour length (log-log plot).}
\label{N15_NX32_AC}
\end{figure}

The pore shape can also be studied via the correlation between the area and the contour length.
Figure \ref{N15_NX32_AC}  shows the pore area $a$ as a function of the contour length $c$ in 
a log-log representation. 
The simulation data are well represented by  the equation
\begin{equation}
\label{scaling_AC}
a\propto c ^{3/2}.
\end{equation}
Such a relationship is consistent with the results obtained by Shillcock \textit{ et al.} \cite{Shillcock_BJ_1998}
for their study of the proliferation of pores in fluid membranes. 
As Shillcock \textit{ et al.}  remark, two dimensional flaccid 
vesicles\cite{Leibler_PRL_87} and two-dimensional ring polymers \cite{Flory_book_53}
show the same behavior as pores in  flaccid membranes.  These different objects may  be modeled  
as closed, self-avoiding,  planar random-walks whose energy depends only on the number of steps.
As a consequence, the correlation between the area $a$ and contour length $c$ observed in our simulations 
[see Eq.~(\ref{scaling_AC})] is consistent with  the simple model of the pore energy given by Eq.~(\ref{energy_pore})
(with $\gamma=0$). We emphasize that the present study and Ref.~~\onlinecite{Shillcock_BJ_1998} are not using the same
types of model. Shillcock \textit{ et al.} simulated pores in membranes  at a mesoscopic length-scale,
with Monte Carlo simulations based on Eq.~(\ref{energy_pore}), whereas in our simulations,
Eq.~(\ref{energy_pore}) emerges from the molecular model. 

\subsection{Dynamics of size fluctuations}
\label{time_evol}
In the previous section, we have shown that
static properties of the pores (size and shape distributions)
are in very good accordance with Eq. (1). 
In the following section, we investigate 
the dynamics of the pore size by
 comparing the simulation results to
a simple stochastic model based on Eq.(1). 
First we determine the time-scale of the pore dynamics.
Second, we present the stochastic model and we
check the relevance of the assumptions of the model 
for our simulation results. Finally, 
the ``shrinking-time''  distribution of the pores 
of the model is fitted to the simulation results
and the resulting parameters are briefly discussed.

\begin{figure}[h!]
\vspace*{0.5cm}
\begin{center}
\includegraphics[width=4.5cm, angle =-90]{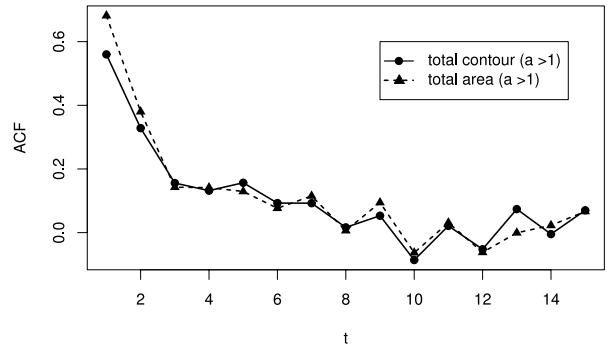}
\caption{Autocorrelation functions of the total areas and contour lengths as a function of time (time unit: $\tau$).
 The pores of area $a=1$ plaquette are not included in the analysis.}
\label{ACF_dt1_contour_area}
\end{center}
\end{figure}

To characterize the short-time dynamics of the contour length $c$, we have measured the 
 number of jumps $\Omega (c,dc)$
 from $c$ to $c+dc$ during one time interval ($\Delta t$).
The probability $P(c,dc)$, that a jump with the initial contour length $c$ has the 
amplitude $dc$ is then defined as
\begin{equation}
P(c,dc) = \frac{\Omega(c,dc)}{\sum_{dc} \Omega(c,dc)}.
\end{equation}
An appropriate numerical value for $\Delta t$ is slightly lower than the typical
time that a pore needs to change its size. The time-scale of pore dynamics 
is expected to be of the same order of magnitude as that
of configurational rearrangements of the amphiphiles, \textit{i.~e.}, 
$10^{-12}\,$s to $10^{-11}\,$s (close to $1\,\tau$). 
To estimate this time-scale, we studied the
correlation time of average quantities like the total contour length, and the total area 
of pore per bilayer (see Fig.~\ref{ACF_dt1_contour_area}).
As expected, the correlation times range around $2\,\tau$.
Given this first estimate of the time-scale dynamics, 
we studied the time-evolution of individual pores every
 $1\,\tau$. 

We compare our results with a  simple stochastic model that has been solved analytically: 
a random walk in a linear potential (RW-LP model~\cite{Khanta_Pramana_83}). 
The RW-LP model describes a random walk in a semi-infinite one-dimensional space with
discrete states labeled $n = 0,1,2,... ,\infty$. For the interpretation of our simulation 
results, the variable $n(t)$ represents one half of the contour length $c(t)$.
The time evolution consist of discrete jumps of amplitude $|dn| =1$, 
at an average rate $W$. The probability to hop towards the upper neighbor ($n \to n+1$) is 
$p^+ = (1-b)/2$.
In the other direction ($n \to n-1$), it is $p^- = (1+b)/2$. The bias $0 \leq  b \leq 1$
measures the tendency to walk towards $n = 0$, where the walker dies (absorbing boundary).
In our interpretation, the parameter $b$ is proportional the line tension of the pores. 
For this model, the probability $Q(n,t|n_0)$  that a walker starting from the state $n_0$
reaches the state $n < n_0$ for the first time after having walked during the time $t$ 
is~\cite{Khanta_Pramana_83}
\begin{eqnarray}
&Q(n,t|n_0) = &\nonumber \\
&\frac{n_0-n}{t} \left(\frac{1+b}{1-b}\right)^{\frac{n_0-n}{2}} e^{-Wt} I_{n_0-n} \left( Wt \sqrt{1-b^2}\right ),&
\label{Q_nt}
\end{eqnarray}
where $I_\nu, (\nu \geq 1)$ is the modified Bessel Function of the first kind, and $W$ and $b$ 
are the two parameters of the model.  In Eq.~(\ref{Q_nt}), the time $t$ is the time needed by the variable $n$ to 
shrink from $n_0 > n$ to $n$, so we call it ``shrinking time" and $Q$ ``shrinking-time distribution". 

Of course, the simulation data are more complicated than the RW-LP model.
(i) The model supposes that  within the time $W^{-1}$, only jumps of amplitude $|dc|=2$ occur.
In the simulations, its is obviously  not the case .
\begin{figure}[h!]
\begin{center}
\includegraphics[width=6cm,angle =-90]{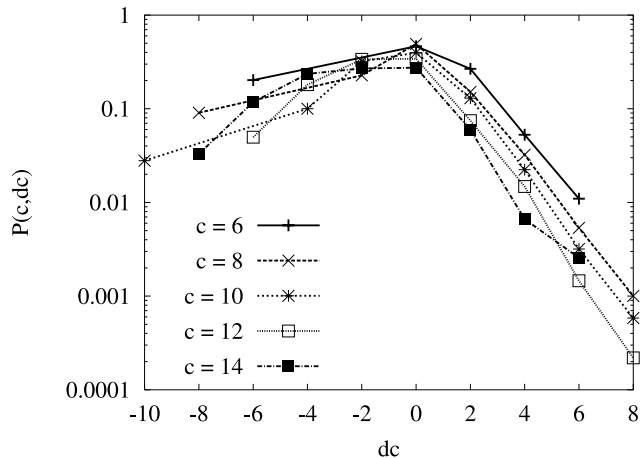}
\caption{Probability of jumps from $c$ to $c+dc$  as a function
 of $dc$ for  possible values of the initial contour length $c$ 
(linear-log scale).}
\label{N5_NX32_dCdt1}
\end{center}
\end{figure}
Nevertheless, the jump probability $P(c,dc)$ decreases when the amplitude  
of the jump  $|dc|$ increases (see Fig.~\ref{N5_NX32_dCdt1}).
For our choice of $\Delta t=1\tau$, 80\%  of the jumps with $dc \neq 0$ correspond to the amplitude $|dc|=2$.
One might expect that if the time of observation is small enough, we should observe
exclusively jumps of the smallest amplitude. We find however that even for the observation 
time $\Delta t  = 0.1\,\tau$, some jumps with large amplitude $|dc| \geq 6$ appear.
The frequency of jumps is thus broadly distributed.

(ii) The model supposes that the effective potential ($-\ln P(c)$) 
is a linear function of $c$. In the simulation, it is not linear
 because of a large entropic contribution to the free energy 
(data not shown here, see Ref~~\onlinecite{Loison_thesis}). Nevertheless,  the effective potential increases
 with the pore size; In  Fig.~\ref{N5_NX32_dCdt1}  the energetic  bias is clearly seen: for a given amplitude $|dc|$,
the probabilities to jump with $dc <0$ are much larger than for $dc >0$.

(iii) The model does not describe the complex behavior of small pores, 
especially the dynamics of hydrophilic pore formation.

\begin{figure}[h!]
\begin{center}
\includegraphics[width=6cm,angle =-90]{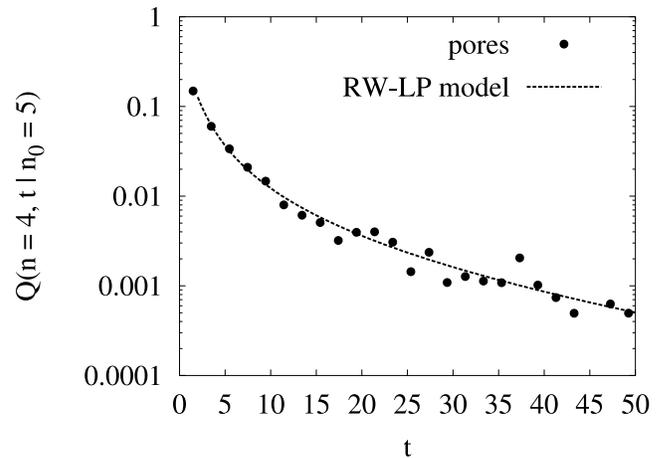}
\caption{Probability that a pore of contour $c = 10$ reaches the contour $c=8$
 for the first time after having fluctuated during the time t (observation time $\Delta t = 1\,\tau$), in a linear-log plot.
The  line represents  Eq.~(\ref{Q_nt}) with  $n_0 = 5, n=4, W = 0.5\,\tau^{-1}$, and $b=0.2$. 
We had to renormalize the pore distribution  to fit it
to Eq.~(\ref{Q_nt}), because it was normalized on the range $t \in [2,450]$
whereas Eq.~(\ref{Q_nt})  is normalized on the range $t \in [0,\infty]$.}
\label{Qt}
\end{center}
\end{figure}

After discussing the assumptions of  the RW-LP model, 
we show that  this stochastic model, yet simple, fit the simulation data. 
We have measured $Q(n,t|n_0)$  for  the simulations using $n_0 =5$ and $n=4$. 
This choice permits to restrict ourselves to large pores ($c \geq 8$);
therefore,  we do not describe the dynamics of the formation of the pores,
but  only  of their shape and size fluctuations.  
As expected from Eq. (\ref{Q_nt}),  other choices of $\{n,n_0\}$ with 
$n_0-n =1$ and $n_0 \geq 5$ give very similar results.
 Fig.~\ref{Qt} shows the comparison between the simulation data (symbols) 
and Eq.~(\ref{Q_nt}) with the parameters $b = 0.2$ and $W = 0.5\,\tau^{-1}$ (curve). 
Despite the simplicity of the model, the agreement is very good and the fitted parameters are reasonable:
The value of the frequency $W=0.5\,\tau^{-1}$ is consistent
 with the previously estimated correlation time of the size fluctuation ($2\,\tau$).
The value of the bias $b=0.2$ is also consistent with the slope $s$ of the free energy 
$-\ln P(c)$ as a function of c ($s=0.1\,k_BT$ per plaquette\cite{footnote2}
To conclude, not only the statics, but also the dynamics of the pore size 
is in good agreement with Lister's model.

\section{Conclusions}
\label{conclusions}

We have investigated a bulk lamellar phase in an amphiphilic
system by molecular dynamics simulations, using a phenomenological
off-lattice model of a binary amphiphile-solvent mixture. The system was 
studied in the (NP$_n$P$_t$T)-ensemble using an extended
Hamiltonian which ensured  that the pressure in the system was isotropic. 
Therefore, the membranes  had no surface tension.
At high amphiphile concentration, (80\% bead percent of amphiphiles), 
the amphiphilic molecules self-assemble into a lamellar phase, \textit{ i.e.},  
a stack of bilayers.  The pores  appearing spontaneously in the bilayers were studied.

The molecular structure of the pores shows
that the  amphiphiles situated in the rim of the pore
reorient: the hydrophilic heads shield the hydrophobic tails from the solvent.

The mean effective free energy of a single pore
was  computed from the distribution of the 
contour-lengths $c$  of the pores. Taking into account the entropic effect of shape fluctuation,
we could fit the simple model $E(c) =E_0+\lambda c$ to the simulation data, and estimate
 the line tension $\lambda$ of the pores. 

Without surface tension, the pores are not circular. 
The relationship between the area $a$  of the pores and their contour-length $c$ is well described by the 
law $a \propto c^{2/3}$.   This relationship was found for other two-dimensional
objects,  whose energy depends only on their contour length  
(models of flaccid vesicles \cite{Leibler_PRL_87} and of self-avoiding ring-polymers\cite{Bishop_JCP_88}).
The analogy between our simulation results and these ``random-walk" models suggest that 
in our analysis, the ``bending energy" of the pore edge is negligible.

Finally, the shrinking-time distribution $Q(n,t|n_0)$  decreases relatively slowly with the time $t$
(it can  be fitted, for example, by the power law ${\cal T}^{-2}$). 
 The  long tail of this probability distribution indicates the existence 
of particularly stable pores.  Presumably, these correspond to pores
 that have grown very large before shrinking again.
Despite its simplicity,  the model of one-dimensional random walk
in a linear potential (RW-LP) \cite{Khanta_Pramana_83} reproduces
 nicely the distribution  of  the  shrinking times observed for the pores. 

To summarize, we have studied in great detail the pore statics
and dynamics in a molecular model, and found that our data are 
in excellent agreement with the predictions of a mesoscopic line 
tension theory $E(c) = E_0 + \lambda c$. 
Our results establish that such simple phenomenological 
models do indeed describe many aspects of real pores in
self-assembled membranes. We have demonstrate that for a 
relatively simple, coarse-grained model, which does however 
treat the particles on a microscopic level. Thus we believe 
that our conclusion will also hold for more realistic models, 
as long as long-range interactions can be neglected.

\begin{acknowledgments}
We thank Kurt Kremer, Peter Reimann, Ralf Eichhorn and Ralf Everaers for fruitful discussions.
We acknowledge the Max Planck Gesellschaft for computation time at the computing center of
Garching.
\end{acknowledgments}

\appendix

\section{Algorithm to detect and analyze the pores.}
\label{appendix_pores}
This appendix describes how we determined the local positions
of membranes in the lamellar stack and the position of the pores in the membranes.
\begin{enumerate}
\item 
The space is divided into $N_xN_yN_z$ cells of size 
$(dx,dy,dz)$ with $N_x=N_y =32$ and $N_z \sim 92$.
For a density of  0.85 particle per volume unit,  
$dx = dy  \simeq 1.3~\sigma$ and $dz \simeq 1.0~\sigma$.
The size of cells may slightly vary from one configuration to another because
the dimensions of the  box  dimensions vary.
\item  
The relative density of tail beads in each cell is calculated
as the ratio $\rho_{tail}(x,y,z) = N_{tail}(x,y,z)/ N_{tot}(x,y,z)$
where $N_{tail}(x,y,z)$ is the number of tail beads in the cell $(x,y,z)$
and $N_{tot}(x,y,z)$ the total number of particles in this same cell. 
\item 
The membranes are defined as the space where the relative density of 
tail beads is higher than a threshold ($\rho_{tail}(x,y,z) > \rho_0$).
The choice of the threshold depends on the mesh size in $x$ and $y$
directions ($dx = {L_x/N_x}$ and $dy = {L_y/N_y}$). 
Typically, we used $\rho_0$ from  0.65 to 0.75 (80 \% of the maximum relative density
of tail beads; values for $dx=dy\sim 1.3 \sigma$).
\item 
The cells that belong to membranes are associated into three dimensional clusters: Two 
membrane-cells that share at least one vortex are attributed to the 
same membrane-cluster. Each membrane-cluster defines a membrane. 
This algorithm identifies membranes even if they have holes. In the presence of necks between
adjacent membranes (local fusion), additional steps have to be taken in order to 
find the necks until one membrane-cluster per membrane is found (not detailed here).
\item 
For each membrane $n$ and each position $(x,y)$, the two heights 
$h^{min}_n(x,y)$ and $h^{max}_n(x,y)$ where the density 
$\rho_{tail}(x,y,z)$ equals the threshold $\rho_0$ are estimated by a 
linear extrapolation. The mean position and the thickness are then
defined by
\begin{eqnarray}
h_n(x,y) & = & \frac{1}{2} \left[  h^{max}_n(x,y) + h^{min}_n(x,y)\right] \nonumber \\
t_n(x,y)  & = &  \frac{1}{2} \left[  h^{max}_n(x,y) - h^{min}_n(x,y)\right]
\label{mean_pos}
\end{eqnarray}
\item The positions $(x,y)$ where the thickness of the membrane $n$ is zero
 or undefined are considered as ``pore-positions". For each membrane,
 an ensemble of  pore-positions $\{ x_i,y_i\}^{pp}_n$ is obtained.
\end{enumerate}

\section{FSA distributions}
We sketch here the principle of our construction of ``finite self-avoiding" (FSA) distribution of
 \verb|NPOINTS| points of size \verb|size| on a grid  of booleans \verb|template| of dimensions \verb|NX*NY|.
Each boolean \verb|template[i,j]| can take the  value \verb|FREE| or \verb|OCCUPIED|.
A schematic algorithm  can be written in the following way 
(not in a real programming language, and without any control about the feasibility of the task !):
\begin{verbatim}
function ConstructFSA(NX,NY,NPOINTS,size)
 n = 0;
 Initialize(template,NX,NY,FREE)
 while(n<NPOINTS)
    xn = NX*random()   	
    yn = NY*random()
    if(template[xn,yn] == FREE) 
          x[n] = xn 
          y[n] = yn 
          Set_occupied(template,xn,yn,size)
          n = n+1 
    end_if	
 end_while
end_function
\end{verbatim}
where \verb|random()| is a random number generator whose output ranges between 0 and 1.
At the end of the loop (it is ever reached !), the vectors \verb|x| and \verb|y| contains the coordinates of the points of the distribution.
The function  \verb|Set_occupied(template,i,j,size)| attributes
the label \verb|OCCUPIED| to all the  sites of the \verb|template| around the site \verb|(i,j)|.
\begin{verbatim}
function Set_occupied(template,i,j,size)
   a = integer((size+1)/2);	
   for(di= -a ; di <= a; di ++)
      for(dj= - a; dj <= a ; dj ++)
         template[i+di,j+dj] = OCCUPIED	
end_function
\end{verbatim}
We have used the following parameters : \verb|NX| = \verb|NY| = 32, \verb|NPOINTS| = 6, \verb|size| = 3, and 
calculated the mean value of the Minkowski functionals over 1000 distributions.

\bibliographystyle{prsty}
\bibliography{paper}

\begin{thebibliography}{10}
\bibitem{footnote1}
The difference between our analysis and the one of Ref.\protect \nobreakspace
  {}\protect \nobreakspace {}\protect \onlinecite {Jacobs_langmuir_98} is that
  our germs are squares, whereas theirs are disks.
\bibitem{footnote2}
To correlate the measurement, we use the (very crude) approximation
  $(1-b)/(1+b)\sim \protect \qopname \relax o{exp}(-4s/k_BT)$ or $1-2b\sim
  1-4s/k_BT$.
\end{thebibliography}

\end{document}